\documentstyle[aps,prc,epsfig]{revtex}
\draft

\newcommand{\ba}{\begin{eqnarray}}
\newcommand{\ea}{\end{eqnarray}}
\def\ii{\'\i}
\begin{document}

\title{Regular spectra in the vibron model 
with random interactions}
\author{R. Bijker$^{1,2}$ and A. Frank$^{2,3}$\\
$^1$ Dipartimento di Fisica, Universit\`a degli Studi di Genova, \\
Via Dodecaneso 33, I-16146 Genova, Italy
\footnote{Sabattical leave} \\
$^2$ Instituto de Ciencias Nucleares, 
Universidad Nacional Aut\'onoma de M\'exico, \\
Apartado Postal 70-543, 
04510 M\'exico, DF, M\'exico \\
$^3$Centro de Ciencias F{\'{\i}}sicas, 
Universidad Nacional Aut\'onoma de M\'exico, \\
Apartado Postal 139-B, 
62251 Cuernavaca, Morelos, M\'exico}

\date{December 18, 2001}

\maketitle

\begin{abstract}
The phenomenom of emerging regular spectral features from random 
interactions is addressed in the context of the vibron model. 
A mean-field analysis links different regions of the 
parameter space with definite geometric shapes. The results that 
are, to a large extent, obtained in closed analytic form, 
provide a clear and transparent interpretation of the high degree 
of order that has been observed in numerical studies. 

\

\end{abstract}

\section{Introduction}

Random matrix theory was developed to describe statistical 
properties of nuclear spectra, such as average distributions and 
fluctuations of peaks in neutron-capture experiments 
\cite{Wigner,Porter}. In this approach, the Hamiltonian matrix 
elements are chosen at random, while keeping some global symmetries, 
e.g. the matrix should be hermitean, and be invariant under 
time-reversal, rotations and reflections. Specific examples include 
the Gaussian Orthogonal Ensemble (GOE) of real-symmetric random 
Hamiltonian matrices in which the many-body interactions are 
uncorrelated, and the Two-Body Random Ensemble (TBRE) in which the 
two-body interactions are taken from a distribution of random 
numbers \cite{Brody}. For two particles, the two ensembles are 
identical but for more than two particles, unlike the case of GOE, 
in TBRE the many-body matrix elements are correlated. As a 
consequence, also the energy eigenvalues of states with different 
quantum numbers are strongly correlated, since they arise from the 
same Hamiltonian. 

The latter aspect was investigated recently in shell model 
calculations for even-even nuclei in the $sd$ shell and 
the $pf$ shell \cite{JBD}. An analysis of the energy spectra of an 
ensemble of random two-body Hamiltonians showed a remarkable 
statistical preference for ground states with angular momentum and 
parity $L^P=0^+$, despite the random nature of the two-body matrix 
elements, both in sign and relative magnitude. A similar 
preponderance of $0^+$ ground states was found in an analysis 
of the Interacting Boson Model (IBM) with random interactions 
\cite{BF1}. In addition, in the IBM evidence was found for both 
vibrational and rotational band structures. According to the 
conventional ideas in the field, the occurrence of $L=0$ ground states 
and the existence of vibrational and rotational bands are due to the 
very specific form of the interactions. Therefore, these unexpected 
and surprising results have sparked a large number of investigations 
to explain and further explore the properties of random nuclei 
\cite{BFP1}-\cite{Zuker}. 

The basic ingredients of the numerical simulations, both for the 
nuclear shell model and for the IBM, are the structure of the 
model space, the ensemble of random Hamiltonians, the order of the 
interactions (one- and two-body), and the global symmetries, i.e. 
time-reversal, hermiticity and rotation and reflection symmetry. 
The latter three symmetries of the Hamiltonian cannot be modified, 
since we are studying many-body systems whose eigenstates have real 
energies and good angular momentum and parity. It has been shown 
that the observed spectral order is a robust property that does 
not depend on the specific  choice of the (two-body) ensemble of 
random interactions \cite{JBD,BFP1,JBDT,DD}, the time-reversal 
symmetry \cite{BFP1}, or the restriction of the Hamiltonian to one- 
and two-body interactions \cite{BF2}. These results suggest that 
that an explanation of the origin of the observed regular features 
has to be sought in the many-body dynamics of the model space 
and/or the general statistical properties of random interactions. 

The purpose of this article is to investigate the origin of the 
regular features that emerge from random interactions in a simple 
exactly solvable model. Hereto we use the vibron model, which is 
mathematically simpler than the IBM, but exhibits many of the same 
qualitative features. A preliminary account of this work has been 
published in \cite{BF3}. 
In Section 2 we present a review of the vibron model by 
studying a schematic Hamiltonian with an arbitrary strength parameter. 
In Section 3 we discuss the spectral properties of an ensemble of 
random one- and two-body interactions, which are interpreted 
in Section 4 in a mean-field analysis. Finally, in Section 5 we 
present our summary and conclusions.

\section{The vibron model}

The vibron model was introduced in 1981 to describe the rotational and 
vibrational excitations of diatomic molecules \cite{vibron}, and has 
also found applications in nuclear cluster models \cite{cluster} and 
meson spectroscopy \cite{meson}. In general terms, it provides an 
algebraic treatment to describe the relative motion in two-body 
problems. The algebraic approach consists in quantizing the relative 
coordinates and momenta with vector boson operators with angular 
momentum and parity $L^P=1^-$ 
\ba
p_{\mu}^{\dagger} &=& \frac{1}{\sqrt{2}}
( r_{\mu} - \frac{\partial}{\partial r_{\mu}} ) ,
\nonumber\\
p_{\mu} &=& \frac{1}{\sqrt{2}}
( r_{\mu} + \frac{\partial}{\partial r_{\mu}} ) ,
\ea
and adding an auxiliary scalar boson with $L^P=0^+$. The 
building blocks of the vibron model are then given by 
\ba
s^{\dagger}, \, p_{\mu}^{\dagger} \hspace{1cm} (\mu=-1,0,1) .
\ea
Sometimes, the scalar and vector bosons are also called vibrons. 
The 16 bilinear products of creation and annihilation operators 
are the generators of the Lie algebra of $U(4)$ 
\ba
s^{\dagger} s, \, s^{\dagger} p_{\mu}, \, p_{\mu}^{\dagger} s, \,
p_{\mu}^{\dagger} p_{\nu} \hspace{1cm} (\mu,\nu=-1,0,1) .
\ea
The Hamiltonian and all other physical operators of interest are 
expressed in terms of the generators. Therefore, the total number 
of vibrons 
\ba
\hat N &=& s^{\dagger} s + \sum_{\mu} p_{\mu}^{\dagger} p_{\mu} ,
\ea
is a conserved quantity. The presence of the scalar boson makes it 
possible to consider, in addition to the three-dimensional harmonic 
oscillator, also situations in which the oscillator shells are mixed. 
In addition to the total number of bosons $N$, the eigenfunctions have 
good angular momentum $L$ and parity $P$.  For a more detailed 
discussion of the vibron model see \cite{book} and references therein. 

\subsection{A schematic Hamiltonian}

For the study of random interactions,  
it is convenient to first consider a schematic Hamiltonian which 
contains the basic features of the model \cite{onno} 
\ba
H &=& - \cos \chi \, p^{\dagger} \cdot \tilde{p} +
\frac{\sin \chi}{4(N-1)} \,
( s^{\dagger} s^{\dagger} - p^{\dagger} \cdot p^{\dagger} ) \,
( \tilde{s} \tilde{s} - \tilde{p} \cdot \tilde{p} ) ,
\label{schematic}
\ea
with $\tilde{s}=s$ and $\tilde{p}_{\mu}=(-1)^{1-\mu} p_{-\mu}$. 
The dots denote a scalar product with respect to the rotation group. 
The range of the angle $\chi$ is that of a full period 
$-\pi/2 < \chi \leq 3\pi/2$, such that all possible combinations of 
attractive and repulsive interactions are included. 

For $\chi=0$, $\pi$ the Hamiltonian has a $U(3)$ dynamic symmetry. 
The spectrum is that of a three-dimensional harmonic oscillator, i.e. 
a series of oscillators shells with 
\ba
n &=& 0,1,\ldots,N ,
\nonumber\\
L &=& n,n-2,\ldots,1 \mbox{ or } 0.
\ea
The parity of the states is $P=(-1)^n=(-1)^L$. 
The energy spectrum is given by 
\ba
E &=& \pm \, n ,
\label{evib}
\ea
where the $+$ $(-)$ sign corresponds to $\chi=0$ $(\pi)$. 

For $\chi=\pi/2$, $3\pi/2$ the Hamiltonian has an $SO(4)$ dynamic 
symmetry. In this case, the harmonic oscillator shells are mixed 
by the Hamiltonian. The spectrum is that of a deformed or Morse 
oscillator, which consists of a series of rotational bands labeled 
by 
\ba
\sigma &=& N,N-2,\ldots,1 \mbox{ or } 0 ,
\nonumber\\
L &=& 0,1,\ldots,\sigma .
\ea
The corresponding energy spectrum is given by 
\ba
E &=& \pm \frac{1}{4(N-1)} (N-\sigma)(N+\sigma+2) 
\nonumber\\
&=& \pm \frac{N+1}{N-1} \, v (1-\frac{v}{N+1}) ,
\ea
where the $+$ $(-)$ sign corresponds to $\chi=\pi/2$ $(3\pi/2)$. 
The ground state band has $v=(N-\sigma)/2=0$ for $\chi=\pi/2$ 
and $v=[N]/2$ for $\chi=3\pi/2$. 

\subsection{Geometric shapes}

\begin{figure}
\centerline{\hbox{\epsfig{figure=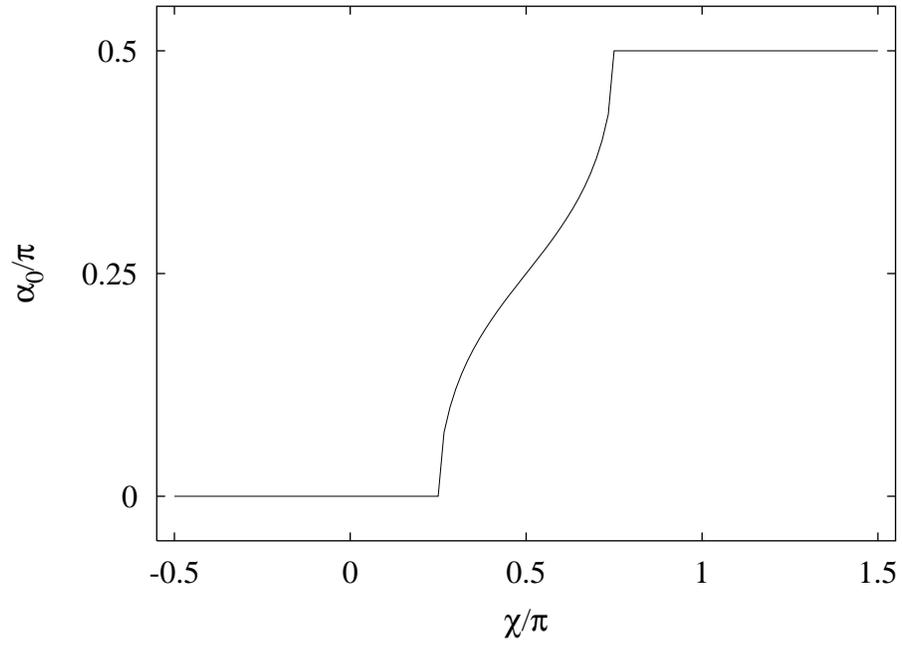} }}
\vspace{15pt}
\caption[]{
Equilibrium configurations of the schematic Hamiltonian of
Eq.~(\protect\ref{schematic}) as a function of $\chi$.}
\label{alfa0}
\end{figure}

The schematic Hamiltonian of Eq.~(\ref{schematic}) exhibits various 
geometric shapes (as well as the phase transitions inbetween them) 
which are relevant for the subsequent studies with random 
interactions. The connection between the vibron model, potential 
energy surfaces, geometric shapes and phase transitions can be 
investigated by means of standard Hartree-Bose mean-field methods 
\cite{onno,ring,duke}. 
For the vibron model, it is convenient to introduce 
a coherent, or intrinsic, state expressed as a condensate of deformed 
bosons with axial symmetry 
\ba
\left| \, N,\alpha \, \right> \;=\; \frac{1}{\sqrt{N!}} \,
\left( \cos \alpha \, s^{\dagger} + \sin \alpha \, p_0^{\dagger}
\right)^N \, \left| \, 0 \, \right> ,
\label{trial}
\ea
with $0 \leq \alpha \leq \pi/2$. The potential energy surface is 
then given by the expectation value of the Hamiltonian in the coherent 
state 
\ba
\frac{1}{N} E(\alpha) &=& \frac{1}{N}
\left< \, N,\alpha \, \right| \, H \, \left| \, N,\alpha \, \right>
\nonumber\\
&=& \cos \chi \, \sin^2 \alpha 
+ \frac{1}{4} \sin \chi \, \cos^2 2\alpha .
\ea
The equilibrium configuration is characterized by the value 
$\alpha=\alpha_0$ for which the potential energy surface attains its 
minimum value 
\ba
\frac{1}{N} \frac{\partial E(\alpha)}{\partial \alpha} \;=\; 0,
&\hspace{1cm}&
\frac{1}{N} \frac{\partial^2 E(\alpha)}{\partial \alpha^2} \;>\; 0.
\ea
The solutions can be divided into three different classes or phases 
\ba
\begin{array}{rclc}
\alpha_0 &=& 0 & -\pi/2 < \chi \leq \pi/4 \\
\cos 2\alpha_0 &=& \cot \chi \hspace{1cm} &\pi/4 \leq \chi \leq 3\pi/4
\\
\alpha_0 &=& \pi/2 & 3\pi/4 \leq \chi \leq 3\pi/2
\end{array}
\ea
which correspond to an $s$-boson or spherical condensate, a deformed 
condensate, and a $p$-boson condensate, respectively (see 
Fig.~\ref{alfa0}). The nature of the phase transitions at the critical 
points $\chi_{\rm c}=\pi/4$, $3\pi/4$ and $3\pi/2$ can be investigated 
by examining the Hartree-Bose ground state energy and its derivates. 
The ground state energy itself is a continuous function of $\chi$
\ba
\frac{1}{N} E(\chi) &=& \left\{ \begin{array}{lll} \frac{1}{4} \sin \chi
&\hspace{1cm}& -\pi/2 < \chi \leq \pi/4 \\
\frac{1}{2} \cos \chi - \frac{\cos^2 \chi}{4\sin \chi}
&\hspace{1cm}& \pi/4 \leq \chi \leq 3\pi/4 \\
\cos \chi + \frac{1}{4} \sin \chi &\hspace{1cm}&
3\pi/4 \leq \chi \leq 3\pi/2 \end{array} \right. 
\label{ehb}
\ea
The first derivative of $E(\chi)$ shows a discontinuity at 
$\chi_{\rm c}=3\pi/2$, and hence the phase transition between the 
spherical, or $s$-boson, condensate and the $p$-boson condensate is of 
first order. The phase transitions involving the deformed condensate 
are of second order, since the second derivative of the ground state 
energy is discontinuous at $\chi_{\rm c}=\pi/4$ and $3\pi/4$.

\subsection{Rotations}

In the previous section, we investigated the equilibrium configurations 
of the schematic Hamiltonian of Eq.~(\ref{schematic}). Each one of them 
corresponds to an intrinsic ground state band $|N,\alpha_0 \rangle$, 
whose angular momentum content depends on 
the value of $\alpha_0$. The rotational energies can be obtained 
by applying standard many-body techniques \cite{ring}. 

In the coherent, or intrinsic, state of Eq.~(\ref{trial}), the 
rotational symmetry is spontaneously broken. In Random Phase 
Approximation the corresponding spurious excitations are decoupled from 
the physical ones and lie at zero energy. The collective or rotational 
energies are then determined by the inertial parameter associated with 
the spurious motion 
\ba
E_{\rm rot} &=& \frac{1}{2{\cal I}} L(L+1) , 
\label{erot}
\ea
where the moment of inertia ${\cal I}$ is obtained from the 
Thouless-Valatin formula. The general procedure is described in 
\cite{ring} and applied to systems of interacting bosons in \cite{duke}. 
For the Hamiltonian of Eq.~(\ref{schematic}), we find that the moment 
of inertia is given by 
\ba
{\cal I} &=& \frac{2N \sin^2 \alpha_0}{\cos 2\alpha_0 \sin \chi} .
\ea
The ordering of the rotational energy levels is then determined 
by the sign of the moment of inertia. In the following, we examine 
each one of the equilibrium configurations in more detail. 

(i) For $\alpha_0=0$, the equilibrium configuration has spherical 
symmetry, and hence can only have $L=0$. The moment of inertia is 
${\cal I}=0$.

(ii) For $0 < \alpha_0 < \pi/2$, the equilibrium shape is deformed. 
The intrinsic state is a condensate of $N$ deformed bosons, which are 
superpositions of monopole and dipole bosons with 
$\cos 2\alpha_0 = \cot \chi$ and $\pi/4 \leq \chi \leq 3\pi/4$. 
The ordering of the rotational energy levels $L=0,1,\ldots,N$ is 
determined by the sign of the moment of inertia 
\ba
{\cal I} &=& \frac{N(\sin \chi - \cos \chi)}{\sin \chi \cos \chi} .
\ea
For $\pi/4 \leq \chi \leq \pi/2$ the moment of inertia is positive 
and hence the ground state has angular momentum $L=0$, whereas for 
for $\pi/2 \leq \chi \leq 3\pi/4$ it is negative corresponding to a 
ground state with $L=N$. 

(iii) For $\alpha_0=\pi/2$, the coherent state is a condensate of $N$ 
dipole or $p$-bosons. The angular momentum content is that of a 
three-dimensional harmonic oscillator shell with $N$ quanta: 
$L=N,N-2,\ldots,1$ or $0$ for $N$ odd or even, respectively. 
The moment of inertia is 
\ba
{\cal I} &=& -\frac{2N}{\sin \chi} .
\ea
This equilibrium shape arises for $3\pi/4 \leq \chi \leq 3\pi/2$. 
For $3\pi/4 \leq \chi \leq \pi$, the moment of inertia is negative 
and the ground state has angular momentum $L=N$. 
For $\pi \leq \chi \leq 3\pi/2$, the moment of inertia is positive and 
the angular momentum of the ground state is $L=0$ for $N$ even and 
$L=1$ for $N$ odd. 

In summary, the schematic Hamiltonian of Eq.~(\ref{schematic}) gives 
rise to three different equilibrium configurations or geometric 
shapes, which are separated by phase transitions. An analysis of the 
angular momentum content of the corresponding condensate combined 
with the sign of the moment of inertia, yields transparent results 
for the ground state angular momentum. The results of 
Table~\ref{gsang} were obtained by assuming a constant probability 
distribution for $\chi$ on the interval $-\pi/2 < \chi \leq 3\pi/2$. 
The ground state is most likely to have angular momentum $L=0$: in 
$75 \%$ of the cases for $N$ even and in $50 \%$ for $N$ odd. 
In $25 \%$ of the cases, the ground state has the maximum value 
of the angular momentum $L=N$. The only other value of the ground 
state angular momentum is $L=1$ in $25 \%$ of the cases for $N$ 
odd. The fluctuation in the $L=0$ and $L=1$ percentages is due to 
the contribution of the $p$-boson condensate. The sum of the $L=0$ 
and $L=1$ percentages is constant ($75 \%$) and does not depend on 
the total number of vibrons $N$. 

An exact analysis in which the Hamiltonian of Eq.~(\ref{schematic}) 
is diagonalized numerically for different values of $\chi$, yields 
identical results for the distribution of the ground state angular 
momenta as are obtained from the mean-field analysis. 

\section{Random interactions}

In this section, we discuss the properties of the vibron model with 
random interactions, or more precisely, with one- and two-body 
interactions with random strengths. We consider the most general one- 
and two-body Hamiltonian of the form 
\ba
H &=& \frac{1}{N} \left[ H_1 + \frac{1}{N-1} H_2 \right] ,
\label{h12}
\ea
where $H_1$ contains the boson energies 
\ba
H_1 &=& \epsilon_s \, s^{\dagger} \cdot \tilde{s}
- \epsilon_p \, p^{\dagger} \cdot \tilde{p} ,
\label{h1}
\ea
and $H_2$ consists of all possible two-body interactions 
\ba
H_2 &=& u_0 \, \frac{1}{2} \,
(s^{\dagger} \times s^{\dagger})^{(0)} \cdot
(\tilde{s} \times \tilde{s})^{(0)}
+ u_1 \, (s^{\dagger} \times p^{\dagger})^{(1)} \cdot
(\tilde{p} \times \tilde{s})^{(1)}
\nonumber\\
&& + \sum_{\lambda=0,2} c_{\lambda} \, \frac{1}{2} \,
(p^{\dagger} \times p^{\dagger})^{(\lambda)} \cdot
(\tilde{p} \times \tilde{p})^{(\lambda)}
\nonumber\\
&& + v_0 \, \frac{1}{2\sqrt{2}} \, \left[
  (p^{\dagger} \times p^{\dagger})^{(0)} \cdot
  (\tilde{s} \times \tilde{s})^{(0)}
+ H.c. \right] .
\label{h2}
\ea
We have scaled $H_1$ by $N$ and $H_2$ by $N(N-1)$ in order to remove 
the $N$ dependence of the matrix elements. The seven parameters of 
the Hamiltonian, altogether denoted by 
\ba
(\vec{x}) &\equiv& (\epsilon_s, \epsilon_p,
u_0, u_1, c_0, c_2, v_0) ,
\ea
are taken as independent random numbers on a Gaussian distribution 
\ba
P(x_i) &=& \mbox{e}^{-x_i^2/2\sigma^2}/\sigma\sqrt{2\pi} ,
\label{gauss}
\ea
with zero mean and width $\sigma$. In this way, the interaction terms 
are arbitrary and equally likely to be attractive or repulsive. The 
spectral properties of each Hamiltonian are analyzed by exact numerical 
diagonalization. The results discussed in this section are based on 
100,000 runs. 

\begin{figure}
\centerline{\hbox{\epsfig{figure=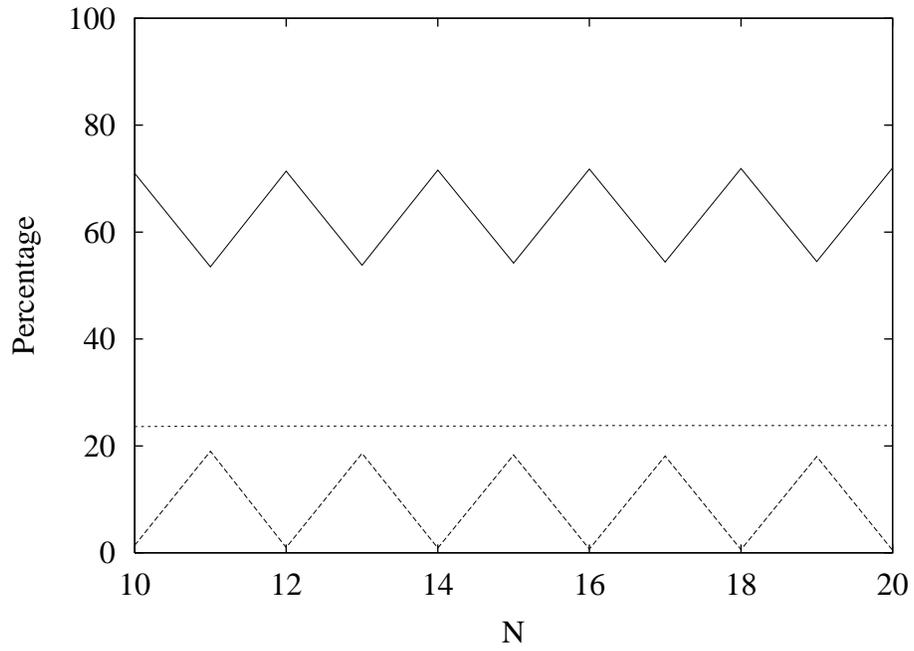} }}
\vspace{15pt}
\caption[]{
Percentage of ground states with angular momentum $L=0$
(solid line), $L=1$ (dashed line) and $L=N$ (dotted line) in the vibron
model with random one- and two-body interactions obtained for
$10 \leq N \leq 20$ and 100000 runs.}
\label{vibgs}
\end{figure}

In Fig.~\ref{vibgs} we show the percentages of $L=0$, $L=1$ and $L=N$ 
ground states as a function of the total number of vibrons $N$. The 
vibron model shows a dominance of $L=0$ ground states. For even values 
of $N$ the ground state has $L=0$ in $\sim 71 \%$ of the cases, and 
for odd values in $\sim 54 \%$ of the cases. Similarly, the 
percentage of ground states with $L=1$ shows an oscillation between 
$\sim 1 \%$ for even values of $N$ and $\sim 18 \%$ for odd values. 
In $\sim 24 \%$ of the cases the ground state has the maximum value 
of the angular momentum $L=N$. 

\begin{figure}
\centerline{\hbox{\epsfig{figure=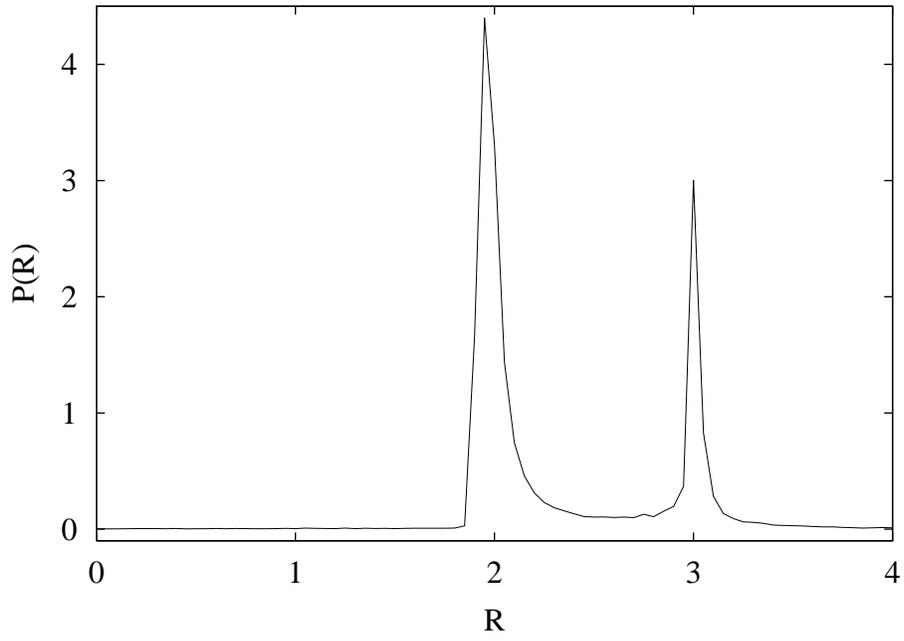} }}
\vspace{15pt}
\caption[]{
Probability distribution $P(R)$ of the energy ratio $R$
of Eq.~(\protect\ref{vibrat}) in the vibron model with random one- and
two-body interactions obtained for $N=19$ and 100000 runs.}
\label{vibpr1}
\end{figure}

\begin{figure}
\centerline{\hbox{\epsfig{figure=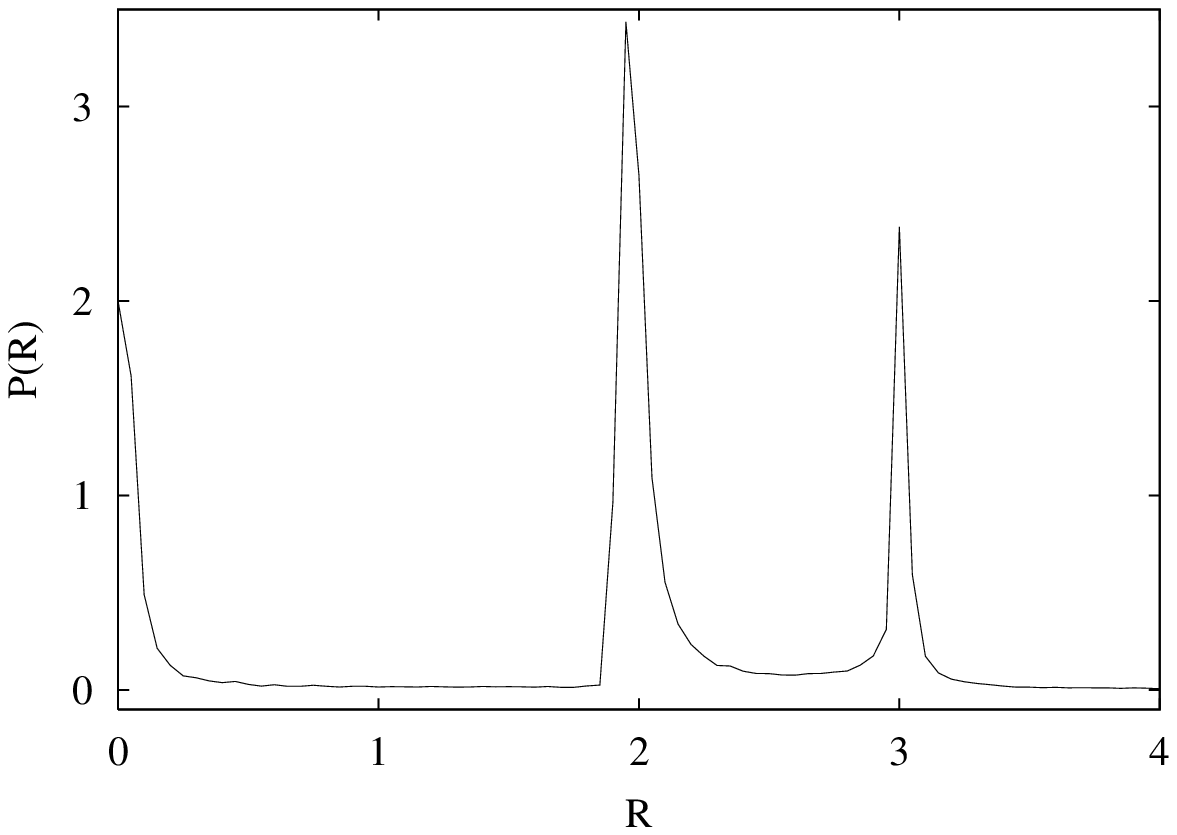} }}
\vspace{15pt}
\caption[]{
As Fig.~\ref{vibpr1}, but for $N=20$.}
\label{vibpr2}
\end{figure}

For the cases with a $L=0$ ground state, it is of interest to study  
the probability distribution of the ratio of excitation energies 
\ba
R &=& \frac{E_{2_1}-E_{0_1}}{E_{1_1}-E_{0_1}} ,
\label{vibrat}
\ea
which constitutes a measure of the spectral properties of the vibron 
Hamiltonian. The energy ratio $R$ has characteristic values of $R=2$ 
for the vibrational or $U(3)$ limit (harmonic oscillator, see 
Eq.~(\ref{evib})), and $R=3$ in the rotational or $SO(4)$ limit 
(Morse oscillator, see Eq.~(\ref{erot})). Figs.~\ref{vibpr1} and 
\ref{vibpr2} show that, both for odd and even values of $N$, the 
probability distribution $P(R)$ has two pronounced peaks, one at the 
vibrational value of $R=2$ and one at the rotational value of $R=3$. 
Moreover, for even values of $N$ there is a maximum at $R=0$, which 
is absent for odd values. 

These numerical results are very similar to those found for the IBM in 
nuclear physics \cite{BF1}, although there are some differences as well. 
Despite the random nature of the interactions both in sign and relative 
magnitude, the spectral properties show a surprisingly large degree of 
order. In recent studies, the tridiagonal form of the Hamiltonian matrix 
in the $U(3)$ basis of the vibron model was used to establish a 
connection with random polynomials \cite{DK}. However, in general the 
Hamiltonian matrix is not of this form, and one has to look for 
alternative methods to understand the origin of these regular properties 
in an analytic and more intuitive way. In the next section, we apply 
the same mean-field techniques that were used in Section~2, to the 
general one- and two-body vibron Hamiltonian of 
Eqs.~(\ref{h12})-(\ref{h2}) with random interaction strengths.

\section{Mean-field analysis}

\begin{figure}
\centerline{\hbox{\epsfig{figure=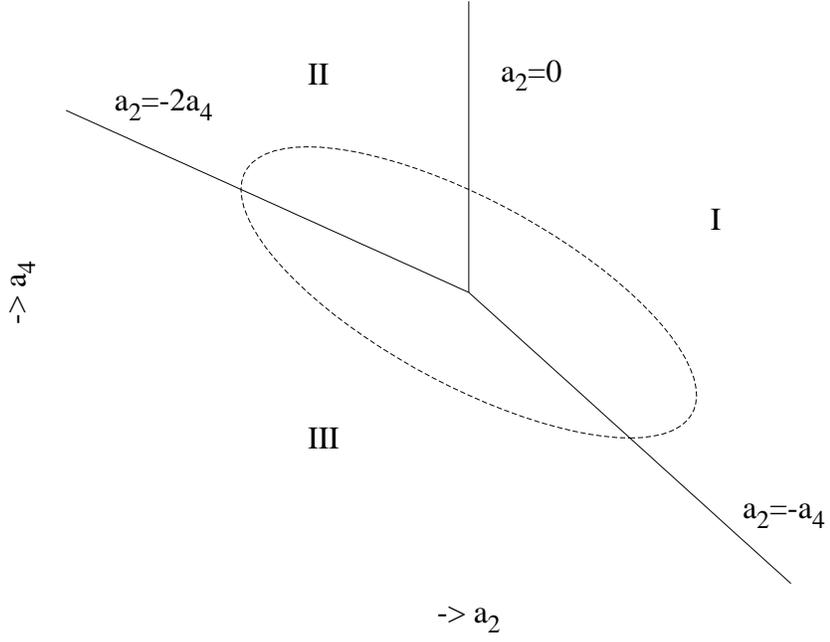} }}
\vspace{15pt}
\caption[]{
Equilibrium configurations in the $a_2 a_4$ plane:
I) spherical shape ($a_2>0$, $a_4>-a_2$), II) deformed shape
($-2a_4<a_2<0$) and III) $p$-boson condensate. The dashed curve
corresponds
to the schematic Hamiltonian of Eq.~(\protect\ref{schematic}), and is
characterized by $a_4=\sin \chi$ and $a_2=\cos \chi - \sin \chi$ with
$\pi/2 < \chi \leq 3\pi/2$.}
\label{a2a4}
\end{figure}

The potential energy surface associated with the general one- and 
two-body vibron Hamiltonian of Eqs.~(\ref{h12})-(\ref{h2}) is given 
by its expectation value in the coherent state of Eq.~(\ref{trial}) 
\ba
E(\alpha) &=&  a_4 \, \sin^4 \alpha + a_2 \, \sin^2 \alpha + a_0 .
\label{surface}
\ea
The coefficients $a_i$ are linear combinations of the parameters 
of the Hamiltonian 
\ba
a_4 &=& \vec{r} \cdot \vec{x} \;=\; \frac{1}{2} u_0 + u_1
+ \frac{1}{6} c_0 + \frac{1}{3} c_2 + \frac{1}{\sqrt{6}} v_0 ,
\nonumber\\
a_2 &=& \vec{s} \cdot \vec{x} \;=\; -\epsilon_s + \epsilon_p
- u_0 - u_1 - \frac{1}{\sqrt{6}} v_0 ,
\nonumber\\
a_0 &=& \vec{t} \cdot \vec{x} \;=\; \epsilon_s + \frac{1}{2} u_0 .
\label{coef}
\ea
For random interaction strengths, the trial wave function of 
Eq.~(\ref{trial}) and the energy surface of Eq.~(\ref{surface}) 
provide information on the distribution of shapes that the model can 
acquire. The value of $\alpha_0$ that characterizes the equilibrium 
configuration of the potential energy surface only depends on the 
coefficients $a_4$ and $a_2$. Just as for the schematic Hamiltonian 
of Eq.~(\ref{schematic}), the parameter space can be divided into 
different areas according to the three possible equilibrium 
configurations 
\ba
\begin{array}{rclc}
\alpha_0 &=& 0 & a_2>0, \;\; a_4>-a2 \\
\sin^2 \alpha_0 &=& -a_2/2a_4 \hspace{1cm} & -2a_4 < a_2 < 0 \\
\alpha_0 &=& \pi/2 & \left\{ \begin{array}{c}
a_2 < 0, \;\; 2a_4 < -a_2 \\ a_4 < -a_2 <0 \end{array} \right.
\end{array}
\ea
In Fig.~\ref{a2a4}, the three regions in the $a_2 a_4$ plane are 
labeled by I for the $s$-boson or spherical condensate ($\alpha_0=0$), 
by II for the deformed condensate ($0 < \alpha_0 < \pi/2$), and by III 
for the $p$-boson condensate ($\alpha_0=\pi/2$). They are separated 
by the separatrices $a_2=0$, $a_4>0$ for I-II, $a_2=-2a_4$, $a_4>0$ for 
II-III, and $a_2=-a_4$, $a_4<0$ for III-I. The dashed curve 
corresponds to the schematic Hamiltonian of Eq.~(\ref{schematic}), and 
is characterized by $a_4=\sin \chi$ and $a_2=\cos \chi - \sin \chi$ 
with $-\pi/2 < \chi \leq 3\pi/2$.  In the previous section, we showed 
that this Hamiltonian exhibits three phase transitions: second 
order transitions at $\chi_{\rm c}=\pi/4$ and $\chi_{\rm c}=3\pi/4$, 
and a first order transition at $\chi_{\rm c}=3\pi/2$. 
The intersections of the dashed 
curve and the separatrices occurs exactly at the critical points 
$\chi_{\rm c}=\pi/4$, $3\pi/4$ and $3\pi/2$. To study the nature of the 
phase transitions for the case of the general Hamiltonian of 
Eqs.~(\ref{h12}-\ref{h2}) we take an arbitrary ellipse in the $a_2 a_4$ 
plane that encloses the origin as its center. It is straightforward to 
show that the order of the phase transitions does not depend on the
orientation nor the eccentricity of the ellipse. In other words, the 
phase transitions are independent of the angle under which the 
separatrices are crossed. 

The distribution of shapes for an ensemble of Hamiltonians depends on 
the joint probability distribution of the coefficients $a_4$ and $a_2$ 
which, for the Gaussian distribution $P(x_i)$ of Eq.~(\ref{gauss}), 
is given by a bivariate normal distribution 
\ba
P(a_4,a_2) &=& \int \prod_{i=1}^7 \, dx_i \, P(x_i) \,
\delta(a_4-\vec{r} \cdot \vec{x}) \, \delta(a_2-\vec{s} \cdot \vec{x})
\nonumber\\
&=& \frac{1}{2\pi \sqrt{\det M}} \, \mbox{exp} \left[-\frac{1}{2}
\left( \begin{array}{cc} a_4 & a_2 \end{array} \right) M^{-1}
\left( \begin{array}{c} a_4 \\ a_2 \end{array} \right) \right] ,
\label{bivariate}
\ea
with 
\ba
M &=& \left( \begin{array}{cc}
\vec{r} \cdot \vec{r} & \vec{r} \cdot \vec{s} \\
\vec{r} \cdot \vec{s} & \vec{s} \cdot \vec{s} \end{array}
\right) \;=\; \frac{1}{18} \left( \begin{array}{rr}
28 & -30 \\ -30 & 75 \end{array} \right) .
\ea
The vectors $\vec{r}$ and $\vec{s}$ are defined in Eq.~(\ref{coef}). 
The probability that the equilibrium shape of an ensemble of 
Hamiltonians is spherical can be obtained by integrating $P(a_4,a_2)$ 
over the appropriate range I ($a_2>0$, $a_4>-a_2$) 
\ba
P_1 &=& \int_{\rm I} \, da_4 \, da_2 \, P(a_4,a_2)
\nonumber\\
&=& \frac{1}{4\pi} \left[ \pi + 2 \arctan \left(
\frac{|\vec{s} \cdot \vec{s} + \vec{r} \cdot \vec{s}|}
{\sqrt{\det M}} \right) \right]
\nonumber\\
&=& \frac{1}{4\pi} \left[ \pi+2 \arctan \sqrt{\frac{27}{16}} \right]
\;=\; 0.396 .
\label{P1}
\ea
Similarly, the probability for the occurrence of a deformed shape can 
be derived by integrating $P(a_4,a_2)$ over the area II ($-2a_4<a_2<0$) 
\ba
P_2 &=& \frac{1}{2\pi} \arctan \left( \frac{2\sqrt{\det M}}
{\vec{s} \cdot \vec{s} + 2\vec{r} \cdot \vec{s}} \right)
\nonumber\\
&=& \frac{1}{2\pi} \arctan \sqrt{\frac{64}{3}} \;=\; 0.216 .
\label{P2}
\ea
Finally, the probability for finding the third solution, 
a $p$-boson condensate, is given by 
\ba
P_3 \;=\; 1-P_1-P_2 \;=\; 0.388 .
\label{P3}
\ea
The angular momentum of the ground state for each of the equilibrium 
configurations can be estimated by evaluating the moment of inertia. 
Just as in Section~2, we adopt the Thouless-Valatin prescription, which 
leads to the formula 
\ba
{\cal I} &=& \frac{2N \sin^2 \alpha_0}{4(N-1)\left[
\frac{1}{2\sqrt{6}}v_0 \cos^2 \alpha_0 - \frac{1}{6}(c_0-c_2)
\sin^2 \alpha_0 \right]} .
\label{tv}
\ea
The moment of inertia depends in a complicated way on the parameters 
in the Hamiltonian, both explicitly as seen in the denominator of 
Eq.~(\ref{tv}) and implicitly through $\alpha_0$, which determines the 
equilibrium configuration. For the schematic Hamiltonian of 
Eq.~(\ref{schematic}), it was possible to obtain a closed expression 
for the moment of inertia, since in this case all properties depend on 
a single parameter $\chi$. However, this is not the case for the 
general one- and two-body Hamiltonian that we are considering here. 
Instead we have to solve the problem numerically. 

\begin{figure}
\centerline{\hbox{\epsfig{figure=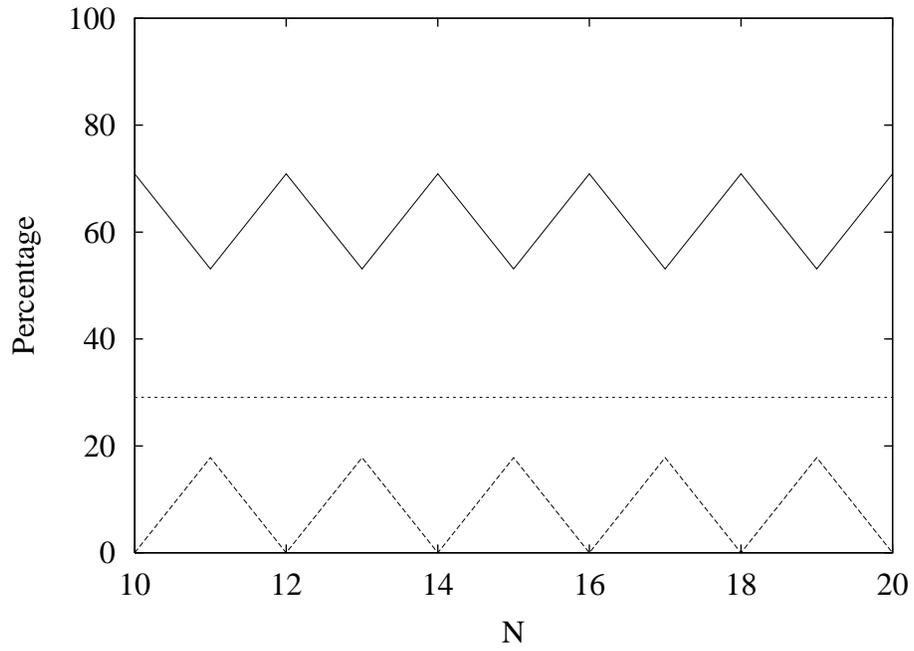} }}
\vspace{15pt}
\caption[]{
Percentage of ground states with angular momentum $L=0$
(solid line), $L=1$ (dashed line) and $L=N$ (dotted line) in the vibron
model with random one- and two-body interactions obtained in a
mean-field
analysis for $10 \leq N \leq 20$.}
\label{vibmf}
\end{figure}

In Table~\ref{gsang1} we show the probability distribution of the 
ground state angular momentum as obtained in the mean-field analysis. 
The results are qualitatively the same as those of Table~\ref{gsang} 
for the schematic Hamiltonian. There is a statistical preference for 
$L=0$ ground states. This is largely due to the occurrence of a 
spherical shape (whose angular momentum content is just $L=0$) for 
almost $40 \%$ of the cases (see Eq.~(\ref{P1})). The deformed shape 
yields ground states either with $L=0$ for positive values of the 
moment of inertia ${\cal I}>0$, or with $L=N$ for ${\cal I}<0$. The 
third solution, the $p$-boson condensate, gives rise to ground states 
with $L=N$ and, depending whether the total number of vibrons is even 
or odd, to $L=0$ or $L=1$, respectively. The sum of the $L=0$ and 
$L=1$ percentages is constant. In Fig.~\ref{vibmf} we show the 
percentages of $L=0$, $L=1$ and $L=N$ ground states, as a function of 
the total number of vibrons $N$. As is clear from the results 
presented in Table~\ref{gsang1}, the fluctuations in the percentages 
of $L=0$ and $L=1$ ground states with $N$ are due to the contribution 
from the $p$-boson condensate solution. A comparison with 
Fig.~\ref{vibgs} shows that the mean-field results are in excellent 
agreement with the exact ones. The difference observed for the $L=N$ 
percentage arises from the fact that in the exact calculations 
for approximately $5 \%$ of the cases, the value of the ground state 
angular momentum is different from $L=0$, $1$, $N$. 

\begin{figure}
\centerline{\hbox{\epsfig{figure=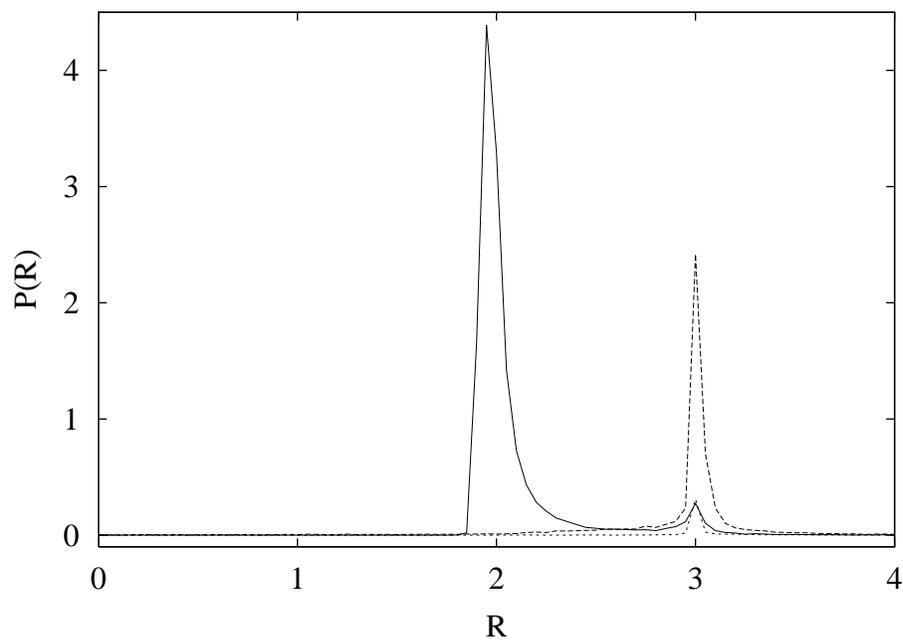} }}
\vspace{15pt}
\caption[]{
Probability distribution $P(R)$ of the energy ratio $R$
obtained for $N=19$ and 100,000 runs
for the spherical (solid line), deformed (dashed line) and $p$-boson
condensate (dotted line) equilibrium configurations, respectively.}
\label{ratio19}
\end{figure}

\begin{figure}
\centerline{\hbox{\epsfig{figure=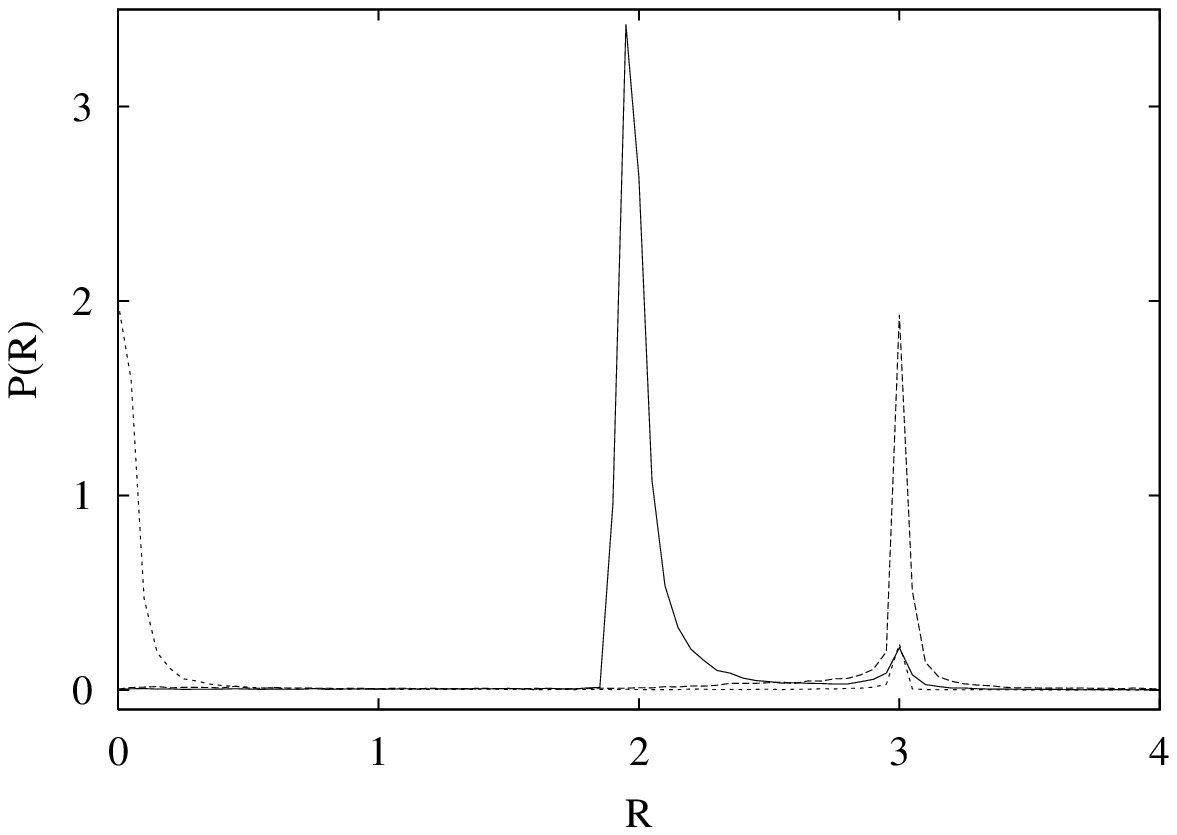} }}
\vspace{15pt}
\caption[]{
As Fig.~\protect\ref{ratio19}, but for $N=20$.}
\label{ratio20}
\end{figure}

In Figs.~\ref{ratio19} and~\ref{ratio20} we show the contribution 
of each one of the equilibrium configurations to the probability 
distribution $P(R)$ of the energy ratio $R$ of Eq.~(\ref{vibrat}) for 
$N=19$ and $N=20$, respectively. For both cases, the spherical shape 
(solid line) contributes almost exclusively to the peak at $R=2$, and 
similarly the deformed shape (dashed line) to the peak at $R=3$, which 
confirms the vibrational and rotational character of these maxima. 
The occurrence of a peak at small values of $R$ for $N=20$ corresponds 
to a level sequence $L=0$, $2$, $1$. It is related to the $p$-boson 
condensate solution (dotted line), which has angular momenta 
$L=N,N-2,\ldots,0$. The first excited $L=1$ state belongs to a 
different band and has a higher excitation energy. For odd values of 
$N$ the $p$-boson condensate has no state with $L=0$, and hence the 
peak at $R=0$ is absent. 

\section{Summary and conclusions}

We have investigated the origin of the regular features that have been 
observed in numerical studies of nuclear structure models with random 
interactions. The observed spectral order is a robust property 
that arises from a much larger class of Hamiltonians than is usually 
thought. It cannot be explained by the time-reversal symmetry of the 
Hamiltonian, the choice of a specific ensemble of random interactions, 
or the restriction to at most two-body interactions. 

In this paper, we have carried out an analysis of the vibron model, 
which is an exactly solvable model to describe the relative motion in 
two-body problems. A numerical study of the vibron model with random 
interactions shows the emergence of ordered features, such as the 
dominance of ground states with $L=0$ and the occurrence of vibrational 
and rotational band structures. In a mean-field analysis, it was found 
that different regions of the parameter space can be associated with 
particular intrinsic vibrational states, which in turn correspond to 
definite geometric shapes: a spherical shape ($\sim 40 \%$), 
a deformed one ($\sim 20 \%$) and a condensate of dipole bosons 
($\sim 40 \%$). Since the spherical shape only has $L=0$, and the 
deformed shape and the $p$-boson condensate with an even number of 
bosons $N$ in about half the number of cases, one finds an $L=0$ ground 
state in approximately $70 \%$ of the cases for $N$ even and $50 \%$ 
for $N$ odd. The spherical shape gives rise to the occurrence of 
vibrational structure, and the deformed shape to rotational bands. 
Qualitatively, these results are very similar to those obtained in 
closed analytic form for a schematic vibron Hamiltonian which 
interpolates between the harmonic oscillator (or $U(3)$ limit) and 
the Morse oscillator (or $SO(4)$ limit). 

In summary, the present results show that a mean-field analysis provides 
a clear and transparent interpretation of the regular features that have 
been obtained in numerical studies of the vibron model with random 
interactions. In \cite{BF3} we have applied similar methods to the IBM. 
Since the structure of the model space of the IBM is more complicated 
than that of the vibron model, the analysis becomes more difficult, but 
the final results are qualitatively the same. The fact that these 
properties are shared by different models, seems to exclude an 
explanation solely in terms of the angular momentum algebra, the 
connectivity of the model space, or the many-body dynamics of the model, 
as has been suggested before. The present analysis points, at least for 
systems of interacting bosons, to a more general phenomenon that 
does not depend so much on the details of the angular momentum coupling, 
but rather on the occurrence of definite, robust geometric phases such 
as spherical and deformed shapes. These shapes are a reflection of 
an intrinsic geometry (or topology) associated to the many-body dynamics 
of the model space which is sampled by the statistical nature of the 
random interactions, but which is quite independent of them. 

For the nuclear shell model the situation is less clear. Although a 
large number of investigations to explain and further explore the 
properties of random nuclei have shed light on various aspects of the 
original problem, i.e. the dominance of $0^+$ ground states, in our 
opinion, no definite answer is yet available, and the full implications 
for nuclear structure physics are still to be clarified.

\section*{Acknowledgments}

This work was supported in part by CONACyT under project Nos. 
32416-E and 32397-E, and by DPAGA under project No. IN106400.

\clearpage

\begin{table}
\centering
\caption[]{
Probabilities of ground states with $L=0$, $1$ and $N$, obtained
in a mean-field analysis of the vibron Hamiltonian of
Eq.~(\protect\ref{schematic}).}
\label{gsang}
\vspace{15pt}
\begin{tabular}{cccccl}
& & & & & \\
\multicolumn{2}{c}{Shape} & $L=0$ & $L=1$ & $L=N$ & \\
& & & & & \\
\hline
& & & & & \\
$  \alpha_0=0$       & 3/8 & 3/8 &  0  &  0  & \\
& & & & & \\
$0< \alpha_0 <\pi/2$ & 1/4 & 1/8 &  0  & 1/8 & \\
& & & & & \\
$  \alpha_0=\pi/2$   & 3/8 & 1/4 &  0  & 1/8 & $N=2k$ \\
                     & 3/8 &  0  & 1/4 & 1/8 & $N=2k+1$ \\
& & & & & \\
\hline
& & & & & \\
Total & 1 & 3/4 &  0  & 1/4 & $N=2k$ \\
      & 1 & 1/2 & 1/4 & 1/4 & $N=2k+1$ \\
& & & & & \\
\end{tabular}
\end{table}

\begin{table}
\centering
\caption[]{
Percentages of ground states with $L=0$, $1$ and $N$, obtained
in a mean-field analysis of the vibron Hamiltonian of
Eqs.~(\protect\ref{h12})-(\protect\ref{h2}).}
\label{gsang1}
\vspace{15pt}
\begin{tabular}{crrrrl}
& & & & & \\
\multicolumn{2}{c}{Shape} & $L=0$ & $L=1$ & $L=N$ & \\
& & & & & \\
\hline
& & & & & \\
$  \alpha_0=0$       & $39.6 \%$
& $39.6 \%$ &  $0.0 \%$ &  $0.0 \%$ & \\
& & & & & \\
$0< \alpha_0 <\pi/2$ & $21.6 \%$
& $13.3 \%$ &  $0.0 \%$ &  $8.3 \%$ & \\
& & & & & \\
$  \alpha_0=\pi/2$   & $38.8 \%$
 & $17.9 \%$ &  $0.0 \%$ & $20.9 \%$ & $N=2k$ \\
&&  $0.0 \%$ & $17.9 \%$ & $20.9 \%$ & $N=2k+1$ \\
& & & & & \\
\hline
& & & & & \\
Total & $100.0 \%$ & $70.8 \%$ &  $0.0 \%$ & $29.2 \%$ & $N=2k$
\\
      &            & $52.9 \%$ & $17.9 \%$ & $29.2 \%$ & $N=2k+1$ \\
& & & & & \\
\end{tabular}
\end{table}

\clearpage

\clearpage

\clearpage

\clearpage

\clearpage

\clearpage

\clearpage

\clearpage

\clearpage

\begin{thebibliography}{aa}

\bibitem{Wigner}
E.P. Wigner,
Ann. of Math. {\bf 62}, 548 (1955);
Ann. of Math. {\bf 67}, 325 (1958).

\bibitem{Porter}
C.E. Porter,
{\em Statistical Theories of Spectra: Fluctuations},
(Academic Press, New York, 1965).

\bibitem{Brody}
T.A. Brody, J. Flores, J.B. French, P.A. Mello, A. Pandey and
S.S.M. Wong, Rev. Mod. Phys. {\bf 53}, 385 (1981); \\
T. Guhr, A.M\"uller-Groeling, H.A. Weidenm\"uller,
Phys. Rep. {\bf 299}, 189 (1998).

\bibitem{JBD}
C.W. Johnson, G.F. Bertsch and D.J. Dean,
Phys. Rev. Lett. {\bf 80}, 2749 (1998).

\bibitem{BF1}
R. Bijker and  A. Frank,
Phys. Rev. Lett. {\bf 84}, (2000), 420.

\bibitem{BFP1}
R. Bijker, A. Frank and S. Pittel,
Phys. Rev. C {\bf 60}, 021302 (1999).

\bibitem{JBDT}
C.W. Johnson, G.F. Bertsch, D.J. Dean and I. Talmi,
Phys. Rev. C {\bf 61}, 014311 (2000).

\bibitem{BF2}
R. Bijker and A. Frank,
Phys. Rev. C {\bf 62}, 014303 (2000).

\bibitem{BFP2}
R. Bijker, A. Frank and S. Pittel,
Rev. Mex. F\ii s. {\bf 46 S1}, 47 (2000).

\bibitem{KZC}
D. Kusnezov, N.V. Zamfir and R.F. Casten,
Phys. Rev. Lett. {\bf 85}, 1396 (2000).

\bibitem{MVZ}
D. Mulhall, A. Volya and V. Zelevinsky,
Phys. Rev. Lett. {\bf 85}, 4016 (2000).

\bibitem{DK}
D. Kusnezov,
Phys. Rev. Lett. {\bf 85}, 3773 (2000); \\
R. Bijker and A. Frank,
Phys. Rev. Lett {\bf 87}, 029201 (2001); \\
D. Kusnezov,
Phys. Rev. Lett {\bf 87}, 029202 (2001).

\bibitem{DW}
S. Dro\.{z}d\.{z} and M. W\'ojcik,
Physica A {\bf 301}, 291 (2001).

\bibitem{ZA}
Y.M. Zhao and A. Arima,
Phys. Rev. C {\bf 64}, 041301 (2001).

\bibitem{DD}
D. Dean,
Nucl. Phys. A {\bf 682}, 194c (2001).

\bibitem{BF3}
R. Bijker and A. Frank,
Phys. Rev. C {\bf 64}, 061303 (2001).

\bibitem{BF4}
R. Bijker and A. Frank,
Nuclear Physics News, Vol. 11, No. 4 (2001), in press.

\bibitem{PvI}
P. Chau Huu-Tai, N.A. Smirnova and P. van Isacker, 
J. Phys. A: Math. Gen., in press.

\bibitem{Zuker}
V. Vel\'azquez and A.P. Zuker,
preprint nucl-th/0106020.

\bibitem{vibron}
F. Iachello,
Chem. Phys. Lett. {\bf 78}, 581 (1981); \\
F. Iachello and R.D. Levine,
J. Chem. Phys. {\bf 77}, 3046 (1982)

\bibitem{cluster}
F. Iachello,
Phys. Rev. C {\bf 23}, 2778 (1981).

\bibitem{meson}
F. Iachello, N.C. Mukhopadhyay and L. Zhang,
Phys. Rev. D {\bf 44},  898 (1991).

\bibitem{book}
F. Iachello and R.D. Levine,
{\it Algebraic Theory of Molecules}
(Oxford University Press, Oxford, 1995).

\bibitem{onno}
O.S. van Roosmalen and A.E.L. Dieperink,
Ann. Phys. (N.Y.) {\bf 139}, 198 (1982); \\
O.S. van Roosmalen, Ph.D. thesis, Univ. of Groningen (1982).

\bibitem{ring}
P. Ring and P. Schuck, 
{\it The Nuclear Many-Body Problem}, 
(Springer Verlag, New York, 1980). 

\bibitem{duke} 
J. Dukelsky, G.G. Dussel, R.P.J. Perazzo, S.L. Reich and
H.M. Sofia,
Nucl. Phys. A {\bf 425}, 93 (1984).

\end{thebibliography}
\end{document}